\documentclass[12pt]{article}

\usepackage{fancyhdr,graphicx,amsmath,amssymb, amsthm}
\usepackage{psfrag}
\usepackage{enumerate}
\usepackage{comment}
\usepackage{url}
\usepackage{amsfonts} 
\usepackage{hyperref} 
\usepackage{bm}

\usepackage{subcaption}
\usepackage[dvipsnames]{xcolor}
\usepackage{float} 
\usepackage{enumitem}
\usepackage{authblk}
\usepackage{pifont}
\usepackage{wrapfig}

\usepackage[top=1.25in, bottom=0.75in, left=1in, right=1in]{geometry}

\usepackage{sectsty}
\newcommand{\spacingset}[1]{%
  \renewcommand{\baselinestretch}{#1}\small\normalsize}

\usepackage[backend=biber,style=apa]{biblatex}
\newtheorem{theorem}{Theorem}

\newtheorem{lemma}[theorem]{Lemma}
\newtheorem{example}{Example}

\newtheorem{assumption}{Assumption}

\DeclareMathOperator{\E}{E}
\DeclareMathOperator{\supp}{supp}
\DeclareMathOperator{\Law}{Law}

\title{Testing Homogeneity of Conditional Distributions: A \(U\)-Statistic Based on Flow Matching}

\author[1]{Changbo Zhu\thanks{Zhu's research was supported by NSF DMS-2412832.}}
\author[2]{Yi Zhao}
\affil[1]{University of Notre Dame, Department of ACMS}
\affil[2]{Indiana University School of Medicine, Department of Biostatistics and Health Data Science}
\date{}

\begin{document}
\spacingset{1}

\maketitle

\begin{abstract}
We propose a flow-matching-based approach for testing the equality of conditional distributions, accommodating both multivariate responses and covariates and allowing the covariate distributions to differ. Our test builds on the key observation that, under suitable conditions, testing the equality of two conditional distributions is equivalent to comparing their corresponding flow-matching vector fields from a common source. The two vector fields agree almost everywhere if and only if the product of two regression residuals,  one for group membership and one for pooled flow matching,  has conditional mean zero given the regression inputs. To test this conditional moment restriction, We use cross-fitted estimates of the residuals to construct a studentized cross \(U\)-statistic that is asymptotically standard normal under the null hypothesis. The proposed test is compared with existing methods in simulations with shifts in the conditional mean or variance and applied to the ACTG 175 trial data.
\end{abstract}

\maketitle

\newpage
\spacingset{1.8}
\section{Introduction}

Testing whether two populations share the same distribution is a fundamental problem. For univariate data, the two-sample Cram\'er--von Mises and Kolmogorov--Smirnov tests compare the empirical distribution functions of the two samples [\cite{anderson1962distribution, smirnov1939estimation}]. For multivariate data, tests based on energy distance [\cite{szekely2004testing}] and maximum mean discrepancy [\cite{gretton2012kernel}] use pairwise distances and kernel evaluations, respectively, without requiring density estimation. Subsequent work has examined the power of these tests in high dimensions [\cite{10.3150/20-BEJ1270, chakraborty2021new}] and adapted energy distance to test equality of pointwise marginal distributions for sparsely observed functional data [\cite{10.1093/jrsssb/qkad021}].

In many applications, however, the question is whether the conditional distribution of a response given covariates is the same across populations, even when the covariate distributions differ. For example, invariant causal prediction uses tests of whether the conditional distribution of a response given a candidate set of covariates remains unchanged across environments to help identify causal covariates. This framework allows the covariate distributions to vary across environments [\cite{peters2016causal}]. In domain adaptation, methods based on covariate shift assume that the conditional response distribution is identical across source and target domains, although their covariate distributions may differ. When response observations are available in both domains, testing equality of conditional distributions provides a way to assess this assumption [\cite{kouw2018introduction}]. Testing equality of conditional distributions also provides a way to evaluate a model by comparing the conditional distribution of model-generated responses with that of observed responses given the same covariates. This comparison arises in conditional model diagnostics using galaxy images [\cite{zhao2021diagnostics}] and in assessing the calibration of probabilistic forecasts based on sequences of satellite images [\cite{dey2022calibrated}].

These applications motivate testing equality of conditional distributions while allowing differences in the underlying covariate distributions. Formally, given a random pair \((Y,X)\) taking values in \(\mathbb{R}^{d_y}\times\mathbb{R}^{d_x}\), let \(P_{Y,X}\) denote its joint distribution, \(P_Y\) and \(P_X\) its marginal distributions, and \(P_{Y\mid X}\) the conditional distribution of \(Y\) given \(X\). Considering two random pairs \((Y^{(1)},X^{(1)})\sim P_{Y^{(1)},X^{(1)}}\) and \((Y^{(2)},X^{(2)})\sim P_{Y^{(2)},X^{(2)}}\) taking values in \(\mathbb{R}^{d_y}\times\mathbb{R}^{d_x}\) and assuming throughout that \(P_{X^{(1)}}\) and \(P_{X^{(2)}}\) admit densities with respect to Lebesgue measure and are mutually absolutely continuous, our goal is to test
\begin{gather*}
H_0:\;
P_{Y^{(1)}\mid X^{(1)}}(\cdot\mid x)
=
P_{Y^{(2)}\mid X^{(2)}}(\cdot\mid x)
\quad\text{for }P_{X^{(1)}}\text{-almost every }x \\
\text{ versus } \\
H_1:\;\text{otherwise}.
\end{gather*}
This problem is challenging for two reasons. Firstly, both the response $Y$ and the covariate $X$ may be multivariate, and $X$ can be high-dimensional.  Secondly, the two samples are generally observed at different covariate values, and the covariate distributions $P_{X^{(1)}}$ and $P_{X^{(2)}}$ need not be the same. Consequently, the joint distributions may differ even when the conditional distributions are identical, so standard two-sample tests applied to the joint observations may reject solely because the covariate distributions differ.

Existing approaches include distance- and kernel-based tests that use kernel smoothing for two independent samples [\cite{yan2022distance}]. However, controlling bias becomes more demanding as the covariate
dimension increases. \cite{chatterjee2024kernel} propose a kernel-based test that uses a nearest-neighbor graph in the covariate space and has a test statistic with mean zero under the null. Their method, however, requires paired responses at shared covariate values, which limits its applicability when the two samples are collected independently. \cite{hu2024two} develop a conformal prediction approach based on a weighted rank-sum statistic. Their method uses an estimated conditional density ratio as the conformity score and marginal covariate density ratios as weights to accommodate differing covariate distributions. These ratios can be estimated through probabilistic classification, avoiding direct density estimation. More recently, \cite{gao2026testing} develop a generative-model-based test that compares observed responses with responses generated from the other population's fitted conditional model at matching covariate values. This avoids conditional density-ratio estimation and local smoothing. Their statistic is a supremum over RKHS function classes, calibrated using a multiplier bootstrap. However, the procedure requires sampling from fitted conditional generators at each observed covariate value and rerunning the alternating maximization separately for every bootstrap replicate. 

\subsection{Our contributions}

We develop a flow-matching framework for testing equality of conditional distributions while allowing the covariate distributions to differ. Flow matching is a generative modeling framework that has attracted considerable attention for its computational efficiency and simple regression-based training. Our main contributions are as follows:
\begin{itemize}
\item We establish that, with a common source distribution and suitable regularity conditions, equality of conditional distributions is equivalent to equality of their associated vector fields. These fields can be conveniently learned through squared-error regression, accommodating multivariate responses and covariates. Our test operates directly on the vector fields, without conditional density estimation, numerical integration of the flow, or sampling from the fitted conditional distributions.

\item We show that the two flow-matching vector fields agree almost everywhere if and only if the product of two regression residuals has conditional mean zero almost surely. These residuals come from regressing the flow-matching target and the group membership indicator on the same inputs under the pooled distribution. To test this conditional moment restriction, we estimate the residuals by cross-fitting and construct a studentized kernel cross $U$-statistic within each evaluation fold following \cite{kim2024dimension}. We combine the fold-specific $p$-values using the Bonferroni correction.

\item Under the null hypothesis and suitable regularity conditions, we show that each studentized fold-specific statistic is asymptotically equivalent to its oracle counterpart and converges in distribution to $N(0,1)$. This result requires both regression estimators to be consistent and the product of their estimation errors to be $o_p(n^{-1/2})$. In particular, it suffices for each estimation error to be $o_p(n^{-1/4})$. The combined test therefore has asymptotic size at most $\alpha$. We also establish consistency against fixed alternatives and give conditions under which power tends to one against local alternatives.
\end{itemize}

\subsection{Notations}
For any
probability distribution $P$, let $\operatorname{supp}(P)$ denote its
support. For two probability measures $P$ and $Q$ on the same measurable space,
we write $P\ll Q$ if $Q(A)=0$ implies $P(A)=0$ for every measurable
set $A$. We say that $P$ and $Q$ are mutually absolutely continuous
if $P\ll Q$ and $Q\ll P$. Following \cite{kosorok2008introduction}, we use $\underset{\mathbf e}{\overset{P}{\rightsquigarrow}}$ to
denote conditional weak convergence in probability with respect
to $\mathbf e$ given remaining data. For two sequences of real numbers $A_n, B_n$, we use $A_n \lesssim B_n$ to mean that there exists a constant $C>0$ such that $A_n \leq C B_n$.

\section{Background on Flow Matching}
Modern generative models such as normalizing flows [\cite{rezende2015variational}], generative adversarial networks (GANs) [\cite{goodfellow2020generative}], and flow matching [\cite{lipman2023flow}] all induce a parametric transport map $f_{\theta}$ that pushes a source distribution to a target distribution. Normalizing flows and GANs models parameterize the transport map $f_{\theta}$ directly, whereas flow matching obtains $f_{\theta}$ implicitly by integrating a learned time-dependent vector field. Among these approaches, flow matching has gained increasing popularity because it learns the vector field through a simple regression objective, avoiding both likelihood maximization required by normalizing flows and adversarial optimization used in GANs.

In particular, let $\rho$ and $P$ denote two probability distributions on $\mathbb{R}^d$, representing the source and target respectively. The goal is to learn a time-dependent vector field \(v(t,z)\) such that the solution to
$$ 
\frac{\mathrm d Z_t}{\mathrm dt}=v(t,Z_t),\qquad Z_0\sim\rho, 
$$
generates a probability path \(Q_t=\Law(Z_t)\) with \(Q_0=\rho\) and \(Q_1=P\). Under suitable regularity conditions, this path satisfies the continuity equation
$
\partial_t Q_t+\nabla\cdot(Q_t v_t)=0$, where we used $v_t(\cdot)=v(t,\cdot)$. Thus, integrating the vector field from \(t=0\) to \(t=1\) transports samples from the source distribution \(\rho\) to the target distribution \(P\).

A vector field transporting \(\rho\) to \(P\) is generally not unique. Flow matching specifies a particular field by taking the conditional mean velocity along the linear interpolation induced by an independent coupling of \(\rho\) and \(P\). To be more specific, let $
Z_0 \sim \rho,
Z_1 \sim P$,
be independent, and define the linear interpolation $Z_t= (1-t)Z_0+t Z_1$. The corresponding \emph{flow-matching vector field} is
\begin{equation}
v(t,z)
=
\E\!\left[
Z_1-Z_0
\,\middle|\,
Z_t = z
\right].
\label{eq:fmvf}
\end{equation}
This vector field induces a probability flow connecting $\rho$ and $P$. 

The flow matching framework extends naturally to conditional distributions. For each fixed $x \in \supp(P_X) $, let $
Z_0\sim \rho, Y\sim P_{Y|X}(\cdot|x)$, be independent, and define the linear interpolation $
Z_t=(1-t)Z_0+tY$. The corresponding \emph{conditional flow-matching vector field} is defined by
\begin{equation}
v(t,z,x)
=
\E\!\left[
Y-Z_0
\,\middle|\,
Z_t=z,\,
X=x
\right].
\label{eq:conditionalfmvf}
\end{equation}
For $P_X$-almost every $x$, the vector field
$v(\cdot,\cdot,x)$ induces a probability flow connecting the common source
distribution $\rho$ to the conditional distribution
$P_{Y|X}(\cdot|x)$. Once the source distribution, interpolation scheme, and coupling are fixed, \emph{the conditional distribution $P_{Y|X}(\cdot | x)$ and the corresponding conditional flow matching vector field are in one-to-one correspondence, up to almost-everywhere versions, under suitable well-posedness conditions}. 

Suppose we observe independent samples $\{(X_i,Y_i)\}_{i=1}^n$. For regression training, generate source samples $Z_{0,1},\ldots,Z_{0,m}\stackrel{\mathrm{i.i.d.}}{\sim}\rho$ and common time points $T_1,\ldots,T_k\stackrel{\mathrm{i.i.d.}}{\sim}\operatorname{Unif}(0,1)$, which we refer to as auxiliary random variables. The source and time samples are independent of each other and of the observations. The conditional flow-matching vector field can be parameterized by a neural
network $v_\theta(t,z,x)$ and estimated by minimizing the regression
loss
\[
\mathcal{L}(\theta)
=
\frac{1}{nmk}
\sum_{i=1}^{n}
\sum_{j=1}^{m}
\sum_{s=1}^{k}
\Bigl\|
v_\theta\!\left(
T_s,
(1-T_s)Z_{0,j}+T_sY_i,
X_i
\right)
-
(Y_i-Z_{0,j})
\Bigr\|_2^2.
\]
The same time point $T_s$ is used for every observation--source
pair $(i,j)$. The population regression target is the conditional flow-matching vector field defined in Equation
\eqref{eq:conditionalfmvf}. Consequently, estimating an entire family of
conditional distributions is reduced to learning a single regression function
$v_\theta(t,z,x)$ over the joint space
$[0,1]\times\mathbb{R}^{d_y}\times\mathbb{R}^{d_x}$.

Compared with classical conditional density estimation, conditional flow matching avoids directly estimating the conditional density $P_{Y|X}(\cdot\mid x)$ for each covariate. Instead, it learns a single regression function $v_\theta(t,z,x)$ over $(t,z,x)$ whose population target is the conditional flow matching vector field. This global regression formulation shares information across all covariates, avoids repeated local smoothing, and scales naturally to high-dimensional covariates. The proposed research builds statistical methodology directly on these learned vector fields, enabling scalable methods.

\section{Methodology}
\label{sec:methodology}

Given two independent samples
$\{(Y_i,X_i)\}_{i=1}^{n_1}
\stackrel{\mathrm{i.i.d.}}{\sim} P_{Y^{(1)},X^{(1)}}$
and
$\{(Y_i,X_i)\}_{i=n_1+1}^{n}
\stackrel{\mathrm{i.i.d.}}{\sim} P_{Y^{(2)},X^{(2)}}$,
we let $n=n_1+n_2$.
Throughout, we choose a source distribution \(\rho\) with a density
that is strictly positive on \(\mathbb{R}^{d_y}\) and has a finite
first moment. The standard Gaussian distribution satisfies these
conditions. Let
$T\sim\operatorname{Unif}(0,1)$ and $Z_0\sim\rho$ be independent
of each other, of the sample $\{(Y_i,X_i)\}_{i=1}^n$,
and of $(Y^{(a)},X^{(a)})$, $a=1,2$. Then, we set
$
Z_t^{(a)}=(1-t)Z_0+tY^{(a)},
W^{(a)}=(T,Z_T^{(a)},X^{(a)})$, $w=(t, z, x)$ and define
the conditional flow matching vector fields as
\[
v^{(a)}(w)=\E[Y^{(a)}-Z_0 \mid
W^{(a)} = w],\qquad a=1,2.
\]
Each vector field $v^{(a)}$ is a regression function with response
$Y^{(a)}-Z_0$ and predictors $W^{(a)}$, and, under suitable regularity
conditions, induces a transport map that transports the common
source $\rho$ to $P_{Y^{(a)}\mid X^{(a)}}(\cdot\mid x)$
for $P_{X^{(a)}}$-almost every $x$. The following theorem formalizes
the characterization of these conditional distributions through
their flow-matching vector fields.

\begin{theorem}
\label{theorem:fm-characterization} 
Assume that, for $P_{X^{(1)}}$-almost every $x$, both conditional
response distributions have finite first moments and
$v^{(a)}(\cdot,\cdot,x)\in
C^1([0,1)\times\mathbb R^{d_y}, \mathbb R^{d_y})$ for $a\in\{1,2\}$. Then the following statements are equivalent:
\begin{enumerate}
\item For
$P_{X^{(1)}}$-almost every $x$,
$P_{Y^{(1)}\mid X^{(1)}}(\cdot\mid x)
=
P_{Y^{(2)}\mid X^{(2)}}(\cdot\mid x)$.

\item
For $P_{X^{(1)}}$-almost every $x$, $
v^{(1)}(t,z,x)=v^{(2)}(t,z,x) \text{ for every }(t,z)\in[0,1)\times\mathbb R^{d_y}$.
\end{enumerate}
\end{theorem}

The $C^1$ condition ensures local existence and uniqueness
of the flow generated by each vector field [\cite[Theorem~1]{lipman2024flow}]. Statement (2) shows that equality
of the conditional distributions is equivalent to equality of
their population flow-matching vector fields.
Parameterizing the two vector fields by $v_{\theta_1}$ and
$v_{\theta_2}$, we can estimate each by applying the flow-matching
regression loss above to its group-specific observations, using
independent auxiliary training draws. Denote the resulting fits by
$\widehat v^{(1)}$ and $\widehat v^{(2)}$.

Then, a natural approach is to compare them through the quadratic plug-in statistic 
$
\sum_{i=1}^n
\left\|
\left(\widehat v^{(1)}-\widehat v^{(2)}\right)
(T_i,Z_{T_i},X_i)
\right\|_2^2
$. However, deriving the asymptotic distribution is challenging, as under null, its leading behavior is determined entirely by the estimation errors of the learned vector fields. Classical empirical process arguments for handling such nuisance estimation typically require strong complexity assumptions on the underlying function class, such as Donsker-type conditions or related entropy restrictions. These assumptions are often too restrictive with modern machine learning estimators, including deep neural networks commonly used to learn flow-matching vector fields. 

To overcome these difficulties, the proposed research reformulates
equality of the two flow-matching vector fields as an equivalent
conditional moment condition. Let $p_1$ and $p_2$ denote the joint
densities of $W^{(1)}$ and $W^{(2)}$, respectively. These densities exist because the covariate distributions and the
source distribution admit Lebesgue densities. Throughout, we assume that $\frac{n_1}{n}\longrightarrow\lambda \in(0,1)$. Define
\begin{itemize}
\item Group label probability: $\pi(w)=\frac{n_1p_1(w)}{n_1p_1(w)+n_2p_2(w)}$.

\item Weighted regression function:
$v(w)=\pi(w)v^{(1)}(w)+\{1-\pi(w)\}v^{(2)}(w)$.

\item Weighted residual:
$
\xi^{(a)}
=\{\mathbb I_{\{a=1\}}-\pi(W^{(a)})\}
  \{(Y^{(a)}-Z_0)-v(W^{(a)})\}.
$
\end{itemize}
If $\E \|\xi^{(a)}\|_2<\infty$ for $a=1,2$, then
for $P_{W^{(1)}}$-almost every $w$, we have $\pi(w)\E[\xi^{(1)}\mid W^{(1)}=w]
+\{1-\pi(w)\} \E [\xi^{(2)}\mid W^{(2)}=w]
=\pi(w)\{1-\pi(w)\}
  \{v^{(1)}(w)-v^{(2)}(w)\}$. The strictly positive source density and mutual absolute continuity
of $P_{X^{(1)}}$ and $P_{X^{(2)}}$ imply mutual absolute continuity
of $P_{W^{(1)}}$ and $P_{W^{(2)}}$. Hence $0<\pi(w)<1$ for
$P_{W^{(1)}}$-almost every $w$. Consequently, under the conditions of
Theorem~\ref{theorem:fm-characterization}, we have $H_0$ holds if and only if for $P_{W^{(1)}}$-almost every $w$,
\begin{equation}
\label{eq:conditionalmean}
\pi(w)\E[\xi^{(1)}\mid W^{(1)}=w]+\{1-\pi(w)\}\E[\xi^{(2)}\mid W^{(2)}=w]=0.
\end{equation}
Hence, the weighted sum of their conditional means is zero
if and only if the two regression functions agree almost everywhere. Here, $\pi$ is the population-level Bayes probabilistic classifier
for group membership under group priors $n_1/n$ and $n_2/n$. The nuisance functions $\pi$ and $v$ are used
to construct the product of residuals, motivated by
Neyman-orthogonal scores in
double machine learning
[\cite{10.1111/ectj.12097}].

Throughout, let $K:([0,1]\times\mathbb R^{d_y+d_x})\times
([0,1]\times\mathbb R^{d_y+d_x})\to\mathbb R$ be a continuous,
symmetric kernel that is integrally strictly positive definite on
finite signed Borel measures
[\cite{sriperumbudur2011universality, muandet2020kernel}], and
satisfies $\sup_w K(w,w)\le\kappa^2<\infty$.
Gaussian and Laplace kernels satisfy these conditions. Let $\mathbb H$ be its RKHS and $\mathbb H^{d_y}$ the
$d_y$-fold product space. Following the kernel conditional moment
testing framework of \cite{muandet2020kernel}, define
\begin{equation}
\label{eq:population-embedding}
\begin{split}
\mu_n
&=\frac{n_1}{n}\E[\xi^{(1)}K(W^{(1)},\cdot)]
+\frac{n_2}{n}\E[\xi^{(2)}K(W^{(2)},\cdot)].
\end{split}
\end{equation}
Under the conditions of Theorem~\ref{theorem:fm-characterization}, Condition
\eqref{eq:conditionalmean} and the conditional-moment embedding
characterization [\cite[Theorem~3.2]{muandet2020kernel}] yield
\[
H_0\text{ holds}
\quad\Longleftrightarrow\quad
\|\mu_n \|_{\mathbb H^{d_y}}^2=0.
\]
This characterization provides a feasible route for testing the null hypothesis. In practice, the nuisance functions $\pi$ and $v$ are unknown and will be
estimated using flexible machine learning methods, including neural networks. To mitigate overfitting and facilitate asymptotic
analysis, we employ stratified $L$-fold cross-fitting with a fixed
integer $L\ge2$. Independently of the observation values, partition
the indices of each sample into $L$ disjoint subsets:
\[
\{1,\ldots,n_1\}
=\bigcup_{\ell=1}^L \mathbb I_\ell^{(1)},
\qquad
\{n_1+1,\ldots,n\}
=\bigcup_{\ell=1}^L \mathbb I_\ell^{(2)},
\]
where, within each sample, the subset sizes differ by at most one.
Combine corresponding subsets to form the evaluation folds $\mathbb I_\ell
=\mathbb I_\ell^{(1)}\cup\mathbb I_\ell^{(2)},
\ell=1,\ldots,L$. 

For each fold $\ell$, estimate $\pi$ and $v$ using observations
outside $\mathbb I_\ell$ and evaluate the residuals on
$\mathbb I_\ell$. For each fold $\ell$,
draw a training source sample $Z_{0,1}^{(-\ell)},\ldots,Z_{0,m_{\mathrm{tr}}}^{(-\ell)}
\stackrel{\mathrm{i.i.d.}}{\sim}\rho$ and a common set of training time points
$
T_1^{(-\ell)},\ldots,T_{k_{\mathrm{tr}}}^{(-\ell)}
\stackrel{\mathrm{i.i.d.}}{\sim}\operatorname{Unif}(0,1),
$
independently of each other and of the observations. For $i\notin\mathbb{I}_\ell$,
define
$
W_{ijs}^{(-\ell)}
=
(
T_s^{(-\ell)},
(1-T_s^{(-\ell)})Z_{0,j}^{(-\ell)}+T_s^{(-\ell)}Y_i,
X_i
)$. We parameterize the two nuisance
functions separately by $\pi_\theta$ and $v_\theta$, and constrain
$\pi_\theta$ to take values in $(0,1)$, for example through a sigmoid
output. Estimate $\pi$ by
minimizing the binary cross-entropy loss
\[
\begin{split}
-\frac{1}{ (n-|\mathbb{I}_\ell|) m_{\mathrm{tr}}k_{\mathrm{tr}}}
\sum_{i\notin\mathbb I_\ell}
\sum_{j=1}^{m_{\mathrm{tr}}}\sum_{s=1}^{k_{\mathrm{tr}}}
\Bigl[
\mathbb I_{\{i\le n_1\}}
\log\pi_\theta(W_{ijs}^{(-\ell)}) +\mathbb I_{\{i>n_1\}}
\log\{1-\pi_\theta(W_{ijs}^{(-\ell)})\}
\Bigr],
\end{split}
\]
and estimate $v$ by minimizing the pooled flow-matching loss
\[
\frac{1}{ (n-|\mathbb{I}_\ell|) m_{\mathrm{tr}}k_{\mathrm{tr}}}
\sum_{i\notin\mathbb{I}_\ell}
\sum_{j=1}^{m_{\mathrm{tr}}}
\sum_{s=1}^{k_{\mathrm{tr}}}
\left\|
v_\theta(W_{ijs}^{(-\ell)})
-\bigl(Y_i-Z_{0,j}^{(-\ell)}\bigr)
\right\|_2^2.
\]
The minimizations are performed separately, i.e.,  $\theta$ denotes the parameters of the corresponding model in each loss.
Denote the resulting estimators by $\widehat\pi^{(-\ell)}$ and
$\widehat v^{(-\ell)}$. Neither estimator uses observations in
$\mathbb{I}_\ell$.

After fitting the nuisance functions, independently for each
observation $i=1,\ldots,n$, generate evaluation source draws
$Z_{0,i1},\ldots,Z_{0,im}\stackrel{\mathrm{i.i.d.}}{\sim}\rho$
and evaluation time points
$T_{i1},\ldots,T_{ik}\stackrel{\mathrm{i.i.d.}}{\sim}
\operatorname{Unif}(0,1)$.
The source and time samples are mutually independent,
independent across observations, and independent of the observed
data, the cross-fitting partition, and all training randomness.
Define $
Z_{ijs}=(1-T_{is})Z_{0,ij}+T_{is}Y_i$ and $
W_{ijs}=(T_{is},Z_{ijs},X_i)$.
For $i\in\mathbb I_\ell$, $j=1,\ldots,m$, and $s=1,\ldots,k$,
define $
\widehat\xi_{ijs}
=
\{\mathbb I_{\{i\le n_1\}}
-\widehat\pi^{(-\ell)}(W_{ijs})\}
\{Y_i-Z_{0,ij}-\widehat v^{(-\ell)}(W_{ijs})\}$,
and
\[
\widehat\Psi_i
=
\frac1{mk}\sum_{j=1}^m\sum_{s=1}^k
\widehat\xi_{ijs}K(W_{ijs},\cdot).
\]
Thus, each residual is evaluated using nuisance estimators trained
outside its evaluation fold. 

We then adopt the sample-splitting and studentization approach
of \cite{kim2024dimension}. For each fold $\ell$, independently partition each
set $\mathbb I_{\ell}^{(a)}$ into two disjoint subsets,
\[
\mathbb I_\ell^{(a)}
=
\mathbb L_\ell^{(a)}\cup\mathbb R_\ell^{(a)},
\qquad a=1,2,
\]
whose sizes differ by at most one. Write
$\mathbb L_\ell=\mathbb L_\ell^{(1)}\cup\mathbb L_\ell^{(2)}$
and
$\mathbb R_\ell=\mathbb R_\ell^{(1)}\cup\mathbb R_\ell^{(2)}$.
Both halves use the same nuisance estimators
$\widehat\pi^{(-\ell)}$ and $\widehat v^{(-\ell)}$,
trained outside $\mathbb I_\ell$.
Using $\mathbb R_\ell$ and the RKHS projection direction $
\frac1{|\mathbb R_\ell|}
\sum_{i'\in\mathbb R_\ell}\widehat\Psi_{i'}$, we  define, for each $i\in\mathbb L_\ell$, the scalar score
\[
\begin{aligned}
\widehat{S}_i
&=
\Bigl\langle
\widehat\Psi_i,\,
\frac1{|\mathbb R_\ell|}
\sum_{i'\in\mathbb R_\ell}\widehat\Psi_{i'}
\Bigr\rangle_{\mathbb H^{d_y}} .
\end{aligned}
\]
The fold-specific cross-$U$ statistic is 
$
\widehat U_\ell
=
\frac1{|\mathbb L_\ell|}
\sum_{i\in\mathbb L_\ell} \widehat{S}_i.
$
For $a=1,2$, define the within-group sample mean and variance by $
\overline{\widehat{S}}_{\ell}^{(a)}
=
\frac1{|\mathbb L_\ell^{(a)}|}
\sum_{i\in\mathbb L_\ell^{(a)}}\widehat{S}_i$ and $
\bigl(\widehat\sigma_\ell^{(a)}\bigr)^2
=
\frac1{|\mathbb L_\ell^{(a)}|-1}
\sum_{i\in\mathbb L_\ell^{(a)}}
\bigl(\widehat{S}_i- \overline{\widehat{S}}_{\ell}^{(a)} \bigr)^2
$ respectively. We define the studentized statistic as
\[
\widehat Z_\ell=\frac{\widehat U_\ell}{\widehat\sigma_\ell},
\]
where $
\widehat\sigma_\ell^2
=
\frac{\sum_{a=1}^2
|\mathbb L_\ell^{(a)}|
\bigl(\widehat\sigma_\ell^{(a)}\bigr)^2}
{|\mathbb L_\ell|^2}$  is the estimated conditional variance of $\widehat U_\ell$. Under the null hypothesis and suitable conditions, $\widehat Z_\ell$ converges in
distribution to a standard normal random variable.
The fold-specific $p$-value is therefore $
p_\ell=1-\Phi(\widehat Z_\ell)$, where $\Phi$ denotes cdf of standard normal distribution. Because observations used for evaluation in one fold can be
used for nuisance training in another, the fold-specific
statistics need not be independent.
We combine their $p$-values using the Bonferroni correction,
\[
p_{\mathrm{Bonf}}
=
\min\left\{1,\;L\min_{1\le\ell\le L} p_\ell \right\},
\]
and reject $H_0$ when $p_{\mathrm{Bonf}}<\alpha$. Provided each fold-specific $p$-value is asymptotically valid,
the combined test has asymptotic size at most $\alpha$.

\section{Asymptotic theory}
\label{sec:asymptotic-theory}

We consider two independent samples of fixed sizes $n_1,n_2$,
with $n=n_1+n_2$ and
$n_1/n\to\lambda_\infty\in(0,1)$.
Throughout this section, the number
of folds $L$, the positive integers $m,k$, and the source
distribution $\rho$ are fixed.
The kernel $K$ is fixed and satisfies the boundedness and
continuity conditions in Section~\ref{sec:methodology}. Define the oracle residuals and features by
\[
\xi_{ijs}
=
\{\mathbb I_{\{i\le n_1\}}-\pi(W_{ijs})\}
\{Y_i-Z_{0,ij}-v(W_{ijs})\},
\qquad
\Psi_i
=
\frac1{mk}\sum_{j=1}^m\sum_{s=1}^k
\xi_{ijs}K(W_{ijs},\cdot).
\]
Hereafter, we omit the subscript $\mathbb H^{d_y}$
from $\|\cdot\|_{\mathbb H^{d_y}}$ and
$\langle\cdot,\cdot\rangle_{\mathbb H^{d_y}}$, using
$\|\cdot\|$ and $\langle\cdot,\cdot\rangle$ for simplicity. At each $n$, the oracle features $\Psi_1,\ldots,\Psi_n$
are mutually independent and identically distributed within
each group. Because  $
\E\Psi_1=\E[\xi^{(1)}K(W^{(1)},\cdot)]$ and $
\E\Psi_{n_1+1}=\E[\xi^{(2)}K(W^{(2)},\cdot)]$, we have $
\frac{n_1}{n}\E\Psi_1
+\frac{n_2}{n}\E\Psi_{n_1+1}
=\mu$. Under $H_0$, it holds that $v^{(1)}=v^{(2)}=v$ almost everywhere, implying
$\E\Psi_i=0$ for every $i$. Then, define the pooled covariance operator
\[
\mathcal C_n h
=
\frac{n_1}{n}\E[\langle\Psi_1,h\rangle\Psi_1]
+
\frac{n_2}{n}
\E[\langle\Psi_{n_1+1},h\rangle\Psi_{n_1+1}],
\qquad h\in\mathbb H^{d_y}.
\]
This operator includes the evaluation auxiliary randomness
and depends on the fixed values of $m,k$.
We denote its trace-norm limit of $\mathcal{C}_n$ by $\mathcal C$. For $i\in\mathbb L_\ell$, define the oracle score
\[
S_i
=
\left\langle
\Psi_i,\,
\frac1{|\mathbb R_\ell|}
\sum_{i'\in\mathbb R_\ell}\Psi_{i'}
\right\rangle.
\]
Let $U_\ell$, $\sigma_\ell$, and $Z_\ell$ be obtained from
$\widehat U_\ell$, $\widehat\sigma_\ell$, and $\widehat Z_\ell$
by replacing the estimated scores $\widehat S_i$ with $S_i$.
In particular, we set
\[
U_\ell
=
\frac1{|\mathbb L_\ell|\,|\mathbb R_\ell|}
\sum_{i\in\mathbb L_\ell}
\sum_{i'\in\mathbb R_\ell}
\langle\Psi_i,\Psi_{i'}\rangle
\text{ and }
Z_\ell=\frac{U_\ell}{\sigma_\ell}.
\]
For definiteness, when a standard error is zero, set the corresponding studentized statistic to $+\infty$ if its
numerator is positive and to zero otherwise.
Under the null conditions below, the probability of a zero
standard error tends to zero.

\begin{assumption}
\label{assumption:asymptotic}
Suppose the following.
\begin{enumerate}[label=(\roman*)]
\item
For some $\eta>0$,
\begin{equation}
\label{eq:m4}
\sup_n\max_{a=1,2}
\E\bigl\|
(Y^{(a)}-Z_0)-v(W^{(a)})
\bigr\|_2^{4+\eta}<\infty.
\end{equation}
Under $H_0$, $\mathcal C_n$ converges in trace norm to a
covariance operator $\mathcal C$ satisfying
$\operatorname{tr}(\mathcal C)>0$.

\item There exist
deterministic sequences $r_{\pi,n},r_{v,n}\to0$ such that
$
\max_{\ell,a}
\|\widehat\pi^{(-\ell)}-\pi\|_{L^4(P_{W^{(a)}})}
=O_p(r_{\pi,n}),
\max_{\ell,a}
\|\widehat v^{(-\ell)}-v\|_{L^4(P_{W^{(a)}})}
=O_p(r_{v,n})$. Moreover,
\begin{equation}
\label{eq:product}
\sqrt n\,r_{\pi,n}r_{v,n}\longrightarrow0.
\end{equation}
Here, for a vector-valued function $f$, the $L^p(P_{W^{(a)}})$-norm is defined as $
\|f\|_{L^p(P_{W^{(a)}})}
=
\{\E\|f(W^{(a)})\|_2^p\}^{1/p}$, where the expectation is over a new independent
$W^{(a)}$, holding $f$ fixed.
\end{enumerate}
\end{assumption}

To test equality of regression functions, \textcite{yan2025machine} fit a separate regression in each group. For comparable group sizes, their Assumption~3 requires each regression estimator, evaluated under the other group's covariate distribution, to be $o_p(n^{-1/4})$. In terms of the convergence rates, our product requirement
\eqref{eq:product} is weaker than requiring both estimation errors
to be $o_p(n^{-1/4})$. For example, if the group-membership
probability is estimated at rate $r_{\pi,n}=O(n^{-1/2})$, then consistency of the pooled flow-matching estimator suffices
for~\eqref{eq:product}.

The moment condition and boundedness of $K$ imply uniformly
bounded $(4+\eta)$th moments of the oracle features.
Under $H_0$ and fixed group laws, $v=v^{(1)}=v^{(2)}$
does not depend on $n$, and the dependence of the oracle
features on $n$ arises only through $\pi$.
Since $n_1/n\to\lambda\in(0,1)$, the group probability
functions converge uniformly, implying $L^2$ convergence
of the oracle features within each group.
Consequently, $\|\mathcal C_n-\mathcal C\|_1\to0$,
where $\|\cdot\|_1$ denotes the trace norm and
$\mathcal C$ is the limiting covariance operator.
The condition $\operatorname{tr}(\mathcal C)>0$ excludes
a degenerate limiting distribution.

\begin{theorem}[Oracle normal limit]
\label{theorem:oracle-limit}
Under Assumption~\ref{assumption:asymptotic} (i) and $H_0$,
for each $\ell=1,\ldots,L$,
\[
Z_\ell\overset{d}{\longrightarrow}N(0,1).
\]
\end{theorem}

The normal limit arises from projecting the features in
$\mathbb L_\ell$ onto the independent empirical feature mean
from $\mathbb R_\ell$ and studentizing the resulting scalar
average. The group-specific variance estimator accommodates
different score distributions in the two groups. We next establish that replacing the oracle nuisance functions
with their estimates preserves the limiting null distribution,
provided that both estimators are consistent and the product of error
rates satisfies $
r_{\pi,n}r_{v,n}=o(n^{-1/2})$.

\begin{theorem}[Studentized oracle equivalence]
\label{theorem:oracle-equivalence}
Under Assumption~\ref{assumption:asymptotic} and $H_0$,
\[
\max_{1\le\ell\le L}
n|\widehat U_\ell-U_\ell|
=o_p(1),
\qquad
\max_{1\le\ell\le L}
\left|
\frac{\widehat\sigma_\ell}{\sigma_\ell}-1
\right|
=o_p(1).
\]
Consequently, $
\max_{1\le\ell\le L}
|\widehat Z_\ell-Z_\ell|
=o_p(1)$ and $
\widehat Z_\ell\overset{d}{\longrightarrow}N(0,1)$.
\end{theorem}

Theorem~\ref{theorem:oracle-equivalence}, together with the
Bonferroni inequality, yields asymptotic size control for
fixed evaluation sample sizes $m$ and $k$, without requiring
independence across the fold-specific statistics.

\begin{theorem}[Asymptotic size]
\label{theorem:size}
Under Assumption~\ref{assumption:asymptotic} and $H_0$,
for every fixed $\alpha\in(0,1)$,
\[
\limsup_{n\to\infty}
P\bigl(p_{\mathrm{Bonf}}<\alpha\bigr)
\le\alpha,
\]
where $
p_{\mathrm{Bonf}}
=
\min  \{
1,\;
L\min_{1\le\ell\le L}
\{1-\Phi(\widehat Z_\ell)\}
 \}$.
\end{theorem}

Having established asymptotic size control, we next examine the power of the proposed test. Allow the conditional distributions
$P_{Y^{(a)}\mid X^{(a)}}$ to dependent on $n$, while keeping the covariate distributions $P_{X^{(a)}}$ fixed, and write
\[
\Delta_n=\|\mu_n\|_{\mathbb H^{d_y}}^2.
\]
The following theorem establishes
consistency when this signal dominates both sampling variability and nuisance-estimation error.

\begin{theorem}[Consistency]
\label{theorem:consistency}
Suppose the moment bound \eqref{eq:m4} and the nuisance
consistency conditions in
Assumption~\ref{assumption:asymptotic}(ii) hold uniformly
along the alternatives, without requiring
\eqref{eq:product}.
If 
\begin{align}
\label{eq:signal}
n\Delta_n\longrightarrow\infty \text{ and }
r_{\pi,n}r_{v,n}=o(\sqrt{\Delta_n}), 
\end{align}
then we have
 $
\max_{1\le\ell\le L}
\left|
\frac{\widehat U_\ell}{\Delta_n}-1
\right|
\overset{p}{\longrightarrow}0$ and, for each $\ell$, $
\widehat Z_\ell\overset{p}{\longrightarrow}+\infty$. Consequently, it holds that
\[
P\bigl(p_{\mathrm{Bonf}}<\alpha\bigr)
\longrightarrow1.
\]
\end{theorem}

The signal conditions \ref{eq:signal} require the population discrepancy to
dominate sampling variation and the product of the nuisance
estimation errors. With independent evaluation draws and
fixed $m,k$, auxiliary randomness contributes to the
observation-level variance and is accounted for by
studentization, rather than appearing as a separate
approximation error. For fixed group laws, $\Delta_n$ converges to the squared
population embedding defined using the limiting group
proportions $\lambda$ and $1-\lambda$.
Under the characterization conditions in
Theorem~\ref{theorem:fm-characterization},
this limit is strictly positive under every fixed alternative.
Together with the moment and nuisance consistency conditions,
this yields consistency for fixed positive integers $m,k$.

\section{Simulation Studies}
\label{sec:empirical-studies}

We compare our method with the mixture density network (MDN) and conditional diffusion model (CDM) implementations of
\cite{gao2026testing} and the linear logistic regression implementation
(HL-LL) of \cite{hu2024two}. 

\subsection{Mean Shift Model}
For a direct comparison, we adopt Model A of
\cite{gao2026testing} with the same data-generating process,
sample sizes, dimensions, and alternative parameters.
All methods are evaluated on the same simulated datasets.
This example introduces a shift in the conditional mean.

\begin{example}[Conditional mean shift]
Let $
\mu=(1,1,-1,-1,0,\ldots,0)^\top\in\mathbb{R}^{d_x}$. Generate independently
\[
X^{(1)}\sim N(0,I_{d_x}),\qquad
X^{(2)}\sim N(\mu,I_{d_x}),\qquad
\varepsilon^{(a)}\sim N(0,I_{d_y}),\quad a=1,2.
\]
For each Monte Carlo repetition, draw a coefficient matrix
$B\in\{-1,1\}^{d_x\times d_y}$ whose entries independently equal
$-1$ or $1$ with probability $1/2$, and use the same matrix
for both groups. Define
\[
Y^{(1)}=B^\top X^{(1)}+\varepsilon^{(1)},\qquad
Y^{(2)}=\delta\mathbf{1}_{d_y}
       +B^\top X^{(2)}+\varepsilon^{(2)},
\]
where $\delta\geq0$ and $\mathbf{1}_{d_y}$ is the vector of ones.
Conditional on $B$, $
Y^{(1)}\mid X^{(1)}=x
\sim N(B^\top x,I_{d_y})$, $ 
Y^{(2)}\mid X^{(2)}=x
\sim N(B^\top x+\delta\mathbf{1}_{d_y},I_{d_y})$. Thus, $\delta=0$ corresponds to the null hypothesis, whereas $\delta>0$ yields different conditional distributions for every
$x\in\mathbb{R}^{d_x}$.
\end{example}

For our method, both nuisance functions are represented by
neural networks with one hidden layer of 256 units and SiLU
activation. Within each training fold, we reserve 20\% of the
observations for validation and use the remaining 80\% for
fitting. Training stops when the validation loss fails to
improve for 30 consecutive epochs or after 600 epochs,
whichever occurs first, with 20 gradient updates per epoch.
We then reinitialize each network and refit it on the entire
training fold, using the selected number of updates and the
learning-rate schedule from the validation stage. Thus,
observations in the evaluation fold are not used for nuisance
training or early stopping. For nuisance training, we use
$m_{\mathrm{train}}=512$ source draws and
$k_{\mathrm{train}}=32$ time points. For evaluation, we independently generate $m=5$ source draws
and $k=2$ uniform time points for each observation. Within
each evaluation fold, we split each group equally into
projection and scoring halves. We use a product Gaussian
kernel with separate bandwidths for the time, interpolated
response, and covariate coordinates. Each bandwidth is
selected using the corresponding median pairwise distance
computed from the projection half only. We compute the
studentized cross-$U$ statistic in each fold and combine
the resulting upper-tail normal $p$-values using the
Bonferroni adjustment. For the MDN and conditional diffusion methods, we use the
network architectures specified in the authors' implementation
of \cite{gao2026testing}.

We consider $(d_x,d_y)\in\{(5,1),(100,1),(100,3)\}$ and
$\delta\in\{0,0.2,0.5,0.8\}$, with $n_1=n_2=2000$ and
500 Monte Carlo repetitions for each setting. All methods
are evaluated on the same generated datasets.
Our method and the methods of \cite{gao2026testing} use
two-fold cross-fitting. The latter use 100 bootstrap
repetitions, whereas our studentized test requires no
bootstrap calibration. We reject the null hypothesis when
the resulting $p$-value is less than $0.05$.
Following \cite{gao2026testing}, we report HL-LL only for
$d_y=1$, as the available implementation is primarily
designed for univariate responses.

Table~\ref{tab:model-a-comparison} reports the empirical
rejection rates. Our method maintains empirical sizes of
$5.2\%$, $5.6\%$, and $5.6\%$, close to the nominal $5\%$
level, and achieves higher empirical power than all
competing methods across the alternatives considered.
For the smaller mean shift $\delta=0.2$, its power is
$64.8\%$, $62.4\%$, and $91.2\%$ in the three dimension
settings, respectively. For $\delta\geq0.5$, its empirical
power reaches $100\%$ in all three settings.

\begin{table}[htbp]
\centering
\caption{Empirical size and power (\%) for Model A at nominal
level $5\%$.}
\label{tab:model-a-comparison}
\begin{tabular}{clrrrr}
\hline
$(d_x,d_y)$ & Method
& $\delta=0$ & $\delta=0.2$ & $\delta=0.5$ & $\delta=0.8$ \\
\hline
$(5,1)$
& Ours            & 5.2 & 64.8 & 100.0 & 100.0 \\
& Gao (diffusion) & 7.2 & 12.0 &  70.2 &  95.8 \\
& Gao (MDN)       & 5.2 &  7.4 &  70.6 &  96.0 \\
& HL-LL           & 5.2 & 45.2 &  98.0 & 100.0 \\
\hline
$(100,1)$
& Ours            & 5.6 & 62.4 & 100.0 & 100.0 \\
& Gao (diffusion) & 2.4 &  3.2 &  16.2 &  75.0 \\
& Gao (MDN)       & 3.8 &  5.4 &  38.2 &  82.0 \\
& HL-LL           & 2.6 & 32.2 &  94.0 & 100.0 \\
\hline
$(100,3)$
& Ours            & 5.6 & 91.2 & 100.0 & 100.0 \\
& Gao (diffusion) & 3.2 &  9.4 &  46.0 &  85.6 \\
& Gao (MDN)       & 6.0 & 10.4 &  54.6 &  80.8 \\
& HL-LL           & N/A & N/A  & N/A   & N/A   \\
\hline
\end{tabular}
\end{table}

\subsection{Variance Shift Model}

We next consider Model B of \cite{gao2026testing}, in which
the alternative changes the conditional variance while
preserving the conditional mean. We adopt the authors' data-generating process and evaluate all methods on the
same simulated datasets.

\begin{example}
Let $B\in\mathbb R^{d_x\times d_y}$ have all entries equal
to $1/\sqrt{d_x}$, and generate
\[
X^{(1)}\sim N(0,I_{d_x}),
\qquad
X^{(2)}\sim N(\mu,I_{d_x}),
\]
where the first $\lfloor d_x/2\rfloor$ coordinates of $\mu$
equal $0.4$ and the remaining coordinates equal $-0.4$.
Define
\[
Y^{(1)}=B^\top X^{(1)}+0.4\varepsilon^{(1)},
\qquad
Y^{(2)}=B^\top X^{(2)}
       +0.4(1+\alpha_2)\varepsilon^{(2)},
\]
where $\alpha_2\geq0$ and
$\varepsilon^{(1)},\varepsilon^{(2)}\sim N(0,I_{d_y})$
are independent of each other and of the covariates.
The two groups have the same conditional mean function,
whereas
$
\operatorname{Var}(Y^{(1)}\mid X^{(1)})
=0.16I_{d_y},
\operatorname{Var}(Y^{(2)}\mid X^{(2)})
=0.16(1+\alpha_2)^2I_{d_y}$. Thus, $\alpha_2=0$ corresponds to the null hypothesis,
and $\alpha_2>0$ introduces a conditional variance shift.
\end{example}

We consider $(d_x,d_y)\in\{(5,1),(100,1),(100,3)\}$ and
$\alpha_2\in\{0,1,2,3,4\}$, with $n_1=n_2=2000$ and
500 Monte Carlo repetitions for each setting.
For our method, both nuisance networks have one hidden
layer of 256 units with ELU activation. All other training,
evaluation, and calibration settings are the same as
those used for Model A, including $m=5$ and $k=2$ with
independent evaluation draws across observations.
For the competing methods, we retain the authors'
implementations and their Model B settings.
As before, HL-LL is reported only for $d_y=1$. Table~\ref{tab:model-b-comparison} reports the empirical
size and power at the nominal $5\%$ level.
Our method maintains empirical sizes between $5.4\%$ and
$6.4\%$ and achieves increasing power as the variance shift
grows despite identical conditional means.
The diffusion method of \cite{gao2026testing} achieves
higher power in the $(100,3)$ setting.
Our approach offers the advantage that studentization yields
an asymptotic standard normal distribution and eliminates
the need for bootstrap calibration.

\begin{table}[htbp]
\centering
\caption{Empirical size and power (\%) for Model B at nominal
level $5\%$.}
\label{tab:model-b-comparison}
\begin{tabular}{clrrrrr}
\hline
$(d_x,d_y)$ & Method
& $\alpha_2=0$ & $\alpha_2=1$ & $\alpha_2=2$
& $\alpha_2=3$ & $\alpha_2=4$ \\
\hline
$(5,1)$
& Ours            & 5.4 & 100.0 & 100.0 & 100.0 & 100.0 \\
& Gao (diffusion) & 4.4 &  55.4 &  98.0 &  99.6 & 100.0 \\
& Gao (MDN)       & 4.4 &  30.6 &  99.8 & 100.0 & 100.0 \\
& HL-LL           & 5.4 &   6.4 &   8.8 &   9.8 &   9.6 \\
\hline
$(100,1)$
& Ours            &  6.4 & 58.4 &  97.0 &  99.8 & 100.0 \\
& Gao (diffusion) &  6.8 & 50.8 & 100.0 & 100.0 & 100.0 \\
& Gao (MDN)       &  3.4 & 26.8 &  99.8 & 100.0 & 100.0 \\
& HL-LL           & 47.4 & 48.0 &  49.4 &  49.6 &  51.0 \\
\hline
$(100,3)$
& Ours            & 6.4 &  56.6 &  96.4 &  99.6 &  99.8 \\
& Gao (diffusion) & 6.0 & 100.0 & 100.0 & 100.0 & 100.0 \\
& Gao (MDN)       & 9.8 &  83.4 & 100.0 & 100.0 & 100.0 \\
& HL-LL           & N/A & N/A   & N/A   & N/A   & N/A   \\
\hline
\end{tabular}
\end{table}

\section{Real Data Experiments}

We apply our test to the publicly available AIDS Clinical
Trials Group Study 175 (ACTG 175) dataset from the UCI
Machine Learning Repository [\cite{kellyUCI}].
ACTG 175 was a randomized clinical trial comparing
antiretroviral therapies in individuals infected with HIV.
The public dataset contains 2,139 participants across four
treatment arms. We consider two arms: zidovudine alone
($n_1=532$) and zidovudine plus didanosine ($n_2=522$).

The goal is to test whether the two treatments yield
different joint distributions of follow-up immune-cell
counts conditional on baseline characteristics.
To this end, we take $X$ to contain CD4 and CD8 counts measured before
treatment together with baseline age and weight.
The response $Y$ consists of CD4 and CD8 counts measured
at $20\pm5$ weeks after treatment initiation. Thus, $d_x=4$ and
$d_y=2$. We test
\[
H_0:\quad
P_{Y^{(1)}\mid X^{(1)}}(\cdot\mid x)
=
P_{Y^{(2)}\mid X^{(2)}}(\cdot\mid x).
\]
Unlike a comparison of conditional means alone, this
hypothesis concerns the entire joint conditional
distribution, including variability and dependence
between the two immune-cell counts. All selected variables are observed for the 1054
participants in these two arms. We retain all these
records and use only baseline measurements as covariates.
For numerical scaling, cell counts are divided by 1,000,
age by 100, and weight by 100. These fixed, invertible
transformations preserve the conditional-distribution
equality hypothesis.

We use two-fold cross-fitting with independent evaluation
draws, setting $m=5$ and $k=2$. Both nuisance networks
have one hidden layer of 256 units with SiLU activation.
We retain the training settings used for the
$(100,3)$ case of Model A, including internal validation
and refitting, without additional hyperparameter tuning.
The product Gaussian kernel uses separate median
bandwidths selected from the projection half.
We fix the random seed at 11. The two fold-specific studentized statistics are
$\widehat Z_1=3.060$ and $\widehat Z_2=2.264$, with
upper-tail $p$-values of $0.00111$ and $0.01179$,
respectively. The Bonferroni-adjusted $p$-value is
$0.00222$, and we therefore reject $H_0$ at the
$5\%$ significance level.
The result provides evidence that the joint conditional
distribution of follow-up CD4 and CD8 counts differs
between the two treatment groups in the public dataset
subset. 

\section{Conclusion}
Flow matching is commonly used to generate samples by integrating a learned vector field. In this work, the vector field itself serves as a statistical representation of a conditional distribution. With a common source distribution and under our regularity conditions, the conditional mean velocity across times and locations characterizes the entire conditional distribution for almost every covariate value. This characterization allows us to test equality of conditional distributions directly through regression residuals. More broadly, this perspective suggests a way to approach other inference problems involving conditional distributions by identifying the restrictions that a hypothesis imposes on the corresponding vector fields.

\printbibliography

\end{document}